\documentclass[%
 aip,
 amsmath,amssymb,
 reprint,%
]{revtex4-1}

\usepackage{graphicx}
\usepackage{dcolumn}
\usepackage{bm}

\usepackage[utf8]{inputenc}
\usepackage[T1]{fontenc}
\usepackage{mathptmx}
\usepackage{etoolbox}
\usepackage{amsmath}
\usepackage{xcolor}

\makeatletter
\def\@email#1#2{%
 \endgroup
 \patchcmd{\titleblock@produce}
  {\frontmatter@RRAPformat}
  {\frontmatter@RRAPformat{\produce@RRAP{*#1\href{mailto:#2}{#2}}}\frontmatter@RRAPformat}
  {}{}
}%
\makeatother
\begin{document}

\preprint{AIP/123-QED}

\title[Article]{High-power even- and odd mode emission from linear arrays of resonant-tunneling-diode (RTD) oscillators in the 0.4- to 0.8-THz frequency range }

\author{Fanqi Meng}
\email{fmeng@physik.uni-frankfurt.de.}
\affiliation{%
Physikalisches Institut, Johann Wolfgang Goethe-Universit\"{a}t, Frankfurt am Main, Germany
}%
\author{Zhenling Tang}
\affiliation{%
Department of Electrical and Electronic Engineering, Tokyo Institute of Technology, Tokyo, Japan
}%
\author{Petr Ourednik}
\affiliation{%
Department of Electrical Engineering and Information Technology, TU Wien, Vienna 1040, Austria
}%
\author{Jahnabi Hazarika}
\affiliation{%
Physikalisches Institut, Johann Wolfgang Goethe-Universit\"{a}t, Frankfurt am Main, Germany
}%
\author{Michael Feiginov}
\affiliation{%
Department of Electrical Engineering and Information Technology, TU Wien, Vienna 1040, Austria
}%
\author{Safumi Suzuki}
\affiliation{%
Department of Electrical and Electronic Engineering, Tokyo Institute of Technology, Tokyo, Japan
}%
\author{Hartmut G. Roskos}
\email{roskos@physik.uni-frankfurt.de.}
\affiliation{%
Physikalisches Institut, Johann Wolfgang Goethe-Universit\"{a}t, Frankfurt am Main, Germany
}%

\date{\today}

\begin{abstract} 
Resonant tunneling diode (RTD) oscillators possess the highest oscillation frequency among all electronic~THz emitters. However, the emitted power from RTDs remains limited. Here, we propose linear RTD-oscillator arrays capable of supporting coherent emission from both odd and even coupled modes. Both modes exhibit constructive interference in the far field, enabling high power emission. Experimental demonstrations of coherent emission from 11-RTD-oscillator linear arrays are presented. The odd mode oscillates at approximately 450~GHz, emitting about 0.5~mW, while the even mode oscillates at around 750~GHz, emitting about 1~mW. Moreover, certain RTD-oscillator arrays demonstrate dual-band oscillation under different biases, allowing for controllable switching between two coupled modes. In addition, during bias sweeping in both directions, a notable hysteresis feature is observed in the switching bias for the odd and even modes. Our linear RTD-oscillator array represents a significant step forward in the realization of high-power large RTD-oscillator arrays and enables large-scale applications of RTD devices.

\end{abstract}

\maketitle
\section{Introduction}

The terahertz (THz) spectrum occupies a unique frequency range between microwave and infrared waves. This spectral region overlaps with both millimeter waves and far-infrared light, offering diverse opportunities for exploration and application. The~THz wave holds significant potential across a myriad of research and practical applications, including non-destructive monitoring/inspection, security, and imaging. Moreover, as the demand for high communication data rates and capacities continues to escalate, wireless communication is increasingly migrating towards~THz frequencies. Despite decades of intensive investigation into~THz technology, a significant gap persists in the development of \textit{high-power, chip-based, and cost-effective}~THz emitters\cite{Tonouchi2007,Asada2021}.

Resonant tunneling diode (RTD) oscillators are room-temperature-operated electronic~THz emitters that have reached, among the electronic~THz emitters, the highest operating frequency \cite{Maekawa2016,Asada2016}. This makes RTD oscillator a prominent candidate for bridging the so-called 'THz gap'. Nevertheless, the reported output power of a single RTD oscillator remains low \cite{Suzuki2013} and the phase noise of the RTD oscillator is also large\cite{Karashima2010}. Substantial efforts have been devoted to the increase of the output power \cite{Feiginov2019}. One natural progression is to aggregate the output power of an array of RTD oscillators through either incoherent power combining or coherent coupling \cite{Asada2008}. This approach not only amplifies the emitted power but also mitigates phase noise \cite{Asada2010}. Recently, there have been several reports on coherent emissions of RTD oscillators. A notable milestone was reached with the coherent coupling of 36 RTD oscillators, resulting in an output power exceeding 10 mW at 0.45~THz \cite{Koyama2022}. Some of the authors of this paper have proposed a planar coherent coupling strategy for slot antennae and verified in experiments that coherent coupling can be achieved via common (shared) stabilization resistors. It was normally assumed that the linear array opts to oscillate in the odd coupled mode, where the phase of the neighboring slot is $\pi$, and the radiation forms destructive interference in the far field. To achieve high power emission, a \textit{zig-zag} structure was proposed to take advantage of the odd mode\cite{VanMai2023} oscillation, and coherent emission from 6 RTD oscillators in the Zig-Zag structure was also reported \cite{Suzuki23}. Additionally, the offset-fed slot antenna was also demonstrated to achieve higher radiation conductance and higher output power. However, understanding the coupled modes among a large number of RTD oscillators within commonly used slot antennae remains a significant challenge. Even mode oscillation of RTD-oscillator in linear slot antenna arrays has not been reported. Substantial progress is still required to achieve high-power emissions for RTD oscillators. Overcoming these challenges will be pivotal in unlocking the full potential of RTD-based THz emitters for a wide range of applications.

In this paper, we propose a novel structure to achieve coherent emission from RTD oscillators integrated into asymmetrically fed linear slot antenna arrays. The array is designed to support both even and odd coupled modes, with the dominance of each mode determined by the mesa area of the RTD: the odd mode prevails at low frequencies for large mesa areas, while the even mode dominates at higher frequencies for smaller mesa areas. Additionally, both odd and even modes exhibit constructive interference in the far field at distinct radiation angles. We fabricated a linear array comprising 11 RTDs and conducted measurements that validated the theoretical predictions. Specifically, when the mesa area of the RTD is large, the array operates in the odd mode, yielding an emission power of approximately 0.5~mW at around 450~GHz. Conversely, for smaller mesa areas, the even mode oscillates at approximately 750~GHz, generating an emission power of about 1~mW. Notably, for certain mesa areas, the RTD array exhibits oscillation at two frequency bands, corresponding to two distinct modes, with bias-controlled switching between them. Moreover, during bidirectional bias sweeping, the switching bias for the odd and even modes displays a hysteresis feature.

\begin{figure}
  \includegraphics[width=0.7\linewidth]{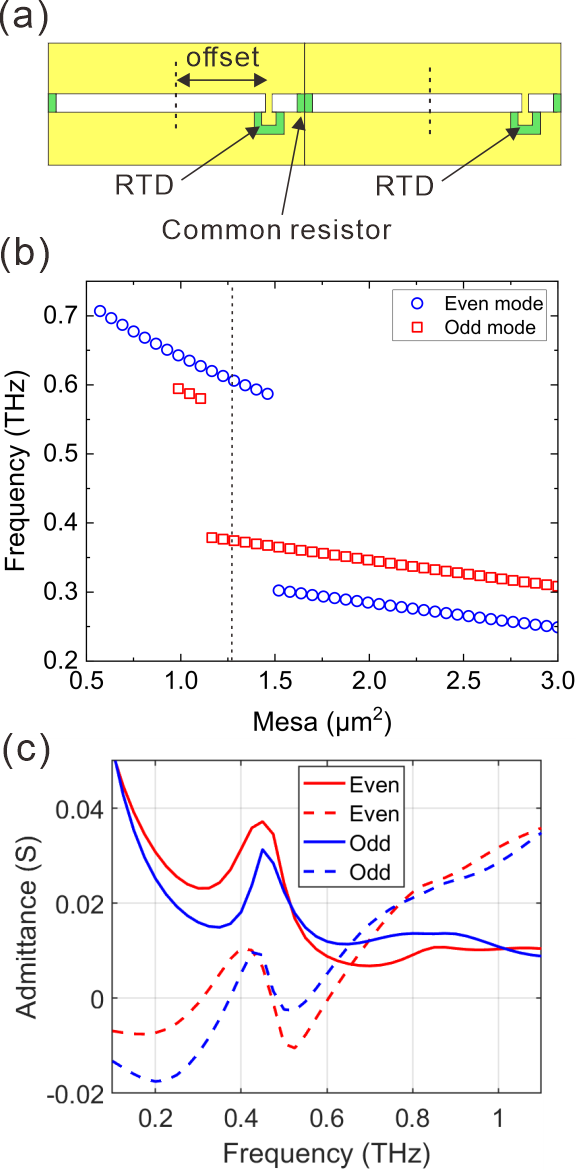}
    \caption{ (a) Sketch of two coupled RTDs integrated with offset-fed slot antennae which are bounded by a common stabilization resistor. (b) Calculated oscillation frequency of the coupled RTDs that are indicated in (a), as a function of mesa area. The blue empty circle represents the even mode and the red empty square represents the odd mode. The vertical dashed line indicates the mesa area (1.3~$\mu$m$^2$) used in the calculations in (c). (c) Calculated admittances of two eigenmodes for a mesa area of 1.3~$\mu$m$^2$. The blue solid/dashed line represents the real/imaginary part of the odd mode, and the red solid/dashed line indicates the real/imaginary part of the even mode.}
  \label{fig:structure}
\end{figure}

\section{Theory and simulation}

The proposed linear RTD oscillator array incorporates an asymmetric offset-fed structure in each slot. The offset-fed slot antenna can increase both oscillation frequency and output power compared with the common centre-fed strategy. Specifically, the short part of the slot offers lower inductance while the long part determines the loop conductance. To analyze the possible oscillation modes in such slot antenna arrays, we simulated a linear array comprising two elements with a common resistor, as illustrated in Fig.~\ref{fig:structure}(a). Each slot is set as 120 $\mu$m in length, with an offset of 47 $\mu$m (20\%) from the slot center, the slot is 5 $\mu$m wide, and the common resistor is 2 $\mu$m in width. We utilize CST electromagnetic solver to derive the admittance properties of the antenna, with RTDs acting as lumped ports. The RTD properties utilized in the simulation are as reported in Ref. \onlinecite{VanTa2022}.
We employ the method proposed in Ref. \onlinecite{VanMai2023} to analyze oscillation modes. By setting the two RTDs as lumped ports, a 2×2 admittance (y) matrix can be derived from the simulation: 
$\begin{pmatrix}
y_{11} & y_{12} \\
y_{21} & y_{22} \\
\end{pmatrix}$, where $y_{12}$ = $y_{21}$.
Diagonalizing the y matrix yields two eigenvalues, as well as two corresponding eigenvectors. The eigenvalues represent the admittance of the two modes, while components of the corresponding eigenvectors denote the phase distribution under that mode. Obviously, at a given frequency, the device tends to oscillate in the mode with the lower eigenvalue real part, as the lower loss makes it easier to be compensated by negative differential conductance (NDC).

Figure~\ref{fig:structure}(b) illustrates the calculated oscillation frequency plotted against the mesa area, showcasing odd (anti-phase) and even (in-phase) modes. In the calculation, the capacitance of RTD is assumed to be 3.3 fF/$\mu m^2$, as verified in the test measurements for this set of RTD wafers. The oscillation frequency demonstrates an increase with decreasing mesa area, attributed to the reduction in the capacitance of the RTD mesa. The blue empty circle denotes the even mode and the red empty square signifies the odd mode. 
There are several salient features: (i) The oscillation frequencies of the even and odd modes lie in two frequency bands: a low-frequency regime below 400~GHz and a high-frequency regime above 550~GHz. (ii) For a mesa area smaller than 1~$\mu$m$^2$, only the even mode oscillates. This occurs at frequencies >0.6~THz in this set of simulations. (iii) For mesas larger than 1~$\mu$m$^2$, both even and odd modes can oscillate. The question arises as to which mode will dominate.

To discern which mode prevails for a large mesa area, Fig.~\ref{fig:structure}(c) depicts the calculated admittances, encompassing both real and imaginary components, of the two eigenmodes when the mesa area is 1.3~$\mu$m$^2$ (corresponds to the dashed line in Fig.~\ref{fig:structure}(b)), serving as an illustrative example. The imaginary parts also contain the contribution of the capacitance of RTDs, i.e. Im ($\gamma$)+$\omega C_{rtd}$, where $C_{rtd}$ is the capacitance of the RTD. When the plotted imaginary part crosses zero at a certain frequency, the pair of RTDs have the potential to oscillate at this frequency, as long as the NDC surpasses the loss in the circuit. The value of loss is mainly determined by the real part of the admittance. In Fig.~\ref{fig:structure}(c), one notices that the odd mode (blue dashed line) oscillates at 0.375~THz. The even mode (red dashed line) exhibits several zero-crossings, one of them at 0.6~THz where the value of the real part of the admittance (i.e., the loss) is lowest. These two oscillation frequencies correspond to the two frequencies that cross the vertical dash line in Fig.~\ref{fig:structure}(b). Since the loss of the even mode (at 0.6~THz ) is smaller than that of the odd mode (at 0.375~THz). The RTD pair would hence preferentially oscillate at 0.6~THz in the even mode. We also calculated the situation when the mesa area is large where both even and odd modes oscillate at the lower frequency band, the odd mode then tends to have a lower loss and hence dominate. 

\begin{figure}
  \includegraphics[width=0.6\linewidth]{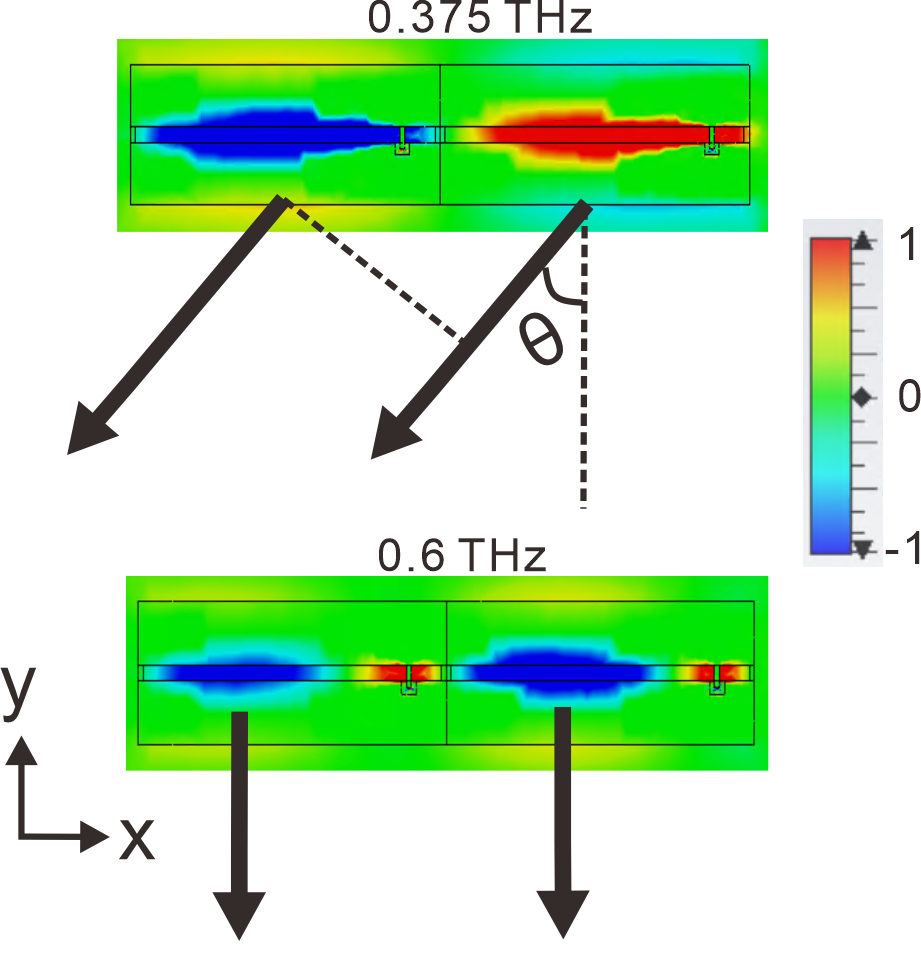}
    \caption{Simulated electric field distributions along y direction (see the coordinate system in the lower left corner).
    Top: For the odd (anti-phase) mode at 0.375~THz, bottom: for the even (in-phase) mode at 0.6~THz. The color bar indicates the electric field strength and field direction at a given moment in time, with red and blue representing opposite directions. 
    Also indicated is the emission direction (the direction for constructive interference at an angle of $\theta$) out through the InP substrate (in the figure, this would be perpendicular to the paper plane; the arrows indicating the directions are rotated for clarity into the paper plane).
    }
  \label{fig:far}
\end{figure}

The prevailing assumption is that the coupled mode possessing a lower loss in the common resistor is expected to dominate the oscillation\cite{VanMai2023}. To be more instructive, Fig.~\ref{fig:far} plots the electric field distribution (along y direction) for the odd mode at 0.375~THz (upper part), and for the even mode at 0.6~THz (bottom part). When the two RTDs oscillate at low frequency (e.g., 0.375~THz), each slot exhibits only one electric field maximum. In the case where the RTD array oscillates at the odd mode, an electric field node forms, resulting in a low alternating current (AC) at the common resistor. As a result, the total loss in the system is reduced. Therefore, the RTD array tends to preferentially oscillate in this mode due to the lower overall loss \cite{VanMai2023}. In contrast, when the RTD array oscillates at a high frequency (e.g., 0.6~THz), each slot typically exhibits two electric field maxima. In this scenario, when the RTD array oscillates at the even mode, an electric field node is formed, resulting in low AC at the common resistor. Then the even mode is preferred to oscillate. This argument qualitatively explains why odd mode dominates at low frequency and even mode is preferred at high frequency for such linear coupled RTD-oscillator arrays.  
\begin{figure}
  \includegraphics[width=0.7\linewidth]{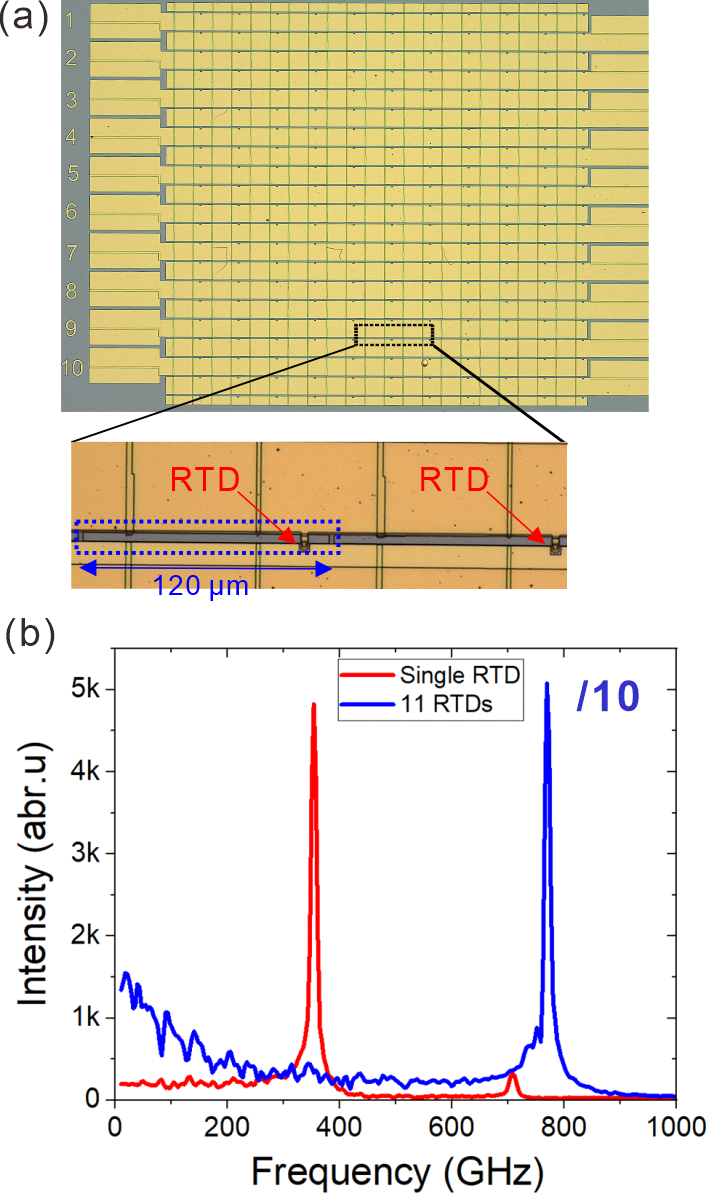}
  \caption{(a) Microscopic image of a fabricated 2D RTD array. Lower panel: Magnified view of two RTDs with offset-fed slot antennas. (b) Emission spectra of a single and a linear RTD array. For better visibility and comparison, the measured intensity of the array was reduced by 10 times. }
  \label{fig:pic}
\end{figure}

For even mode oscillation, since the phase of all the oscillators are the same, constructive interference points perpendicular to the substrate, as indicated by the arrows in the lower part of Fig.~\ref{fig:far}. For odd mode oscillation, the two RTD oscillators possess a $\pi$ phase difference. The constructive interference is only possible at a radiation angle of $\theta$, where the optical path difference induces another phase difference of $\pi$, as shown in the upper part of Fig.~\ref{fig:far}. Assuming the length of the slot, the wavelength, the refractive index of the substrate are $L$,  $\lambda$ and $n$, the constructive interference angle satisfies:
\begin{equation}
     sin(\theta)=\frac{\lambda}{2nL}
\label{equ:phase}
\end{equation}
If the oscillation frequency of the odd mode is larger than 0.37~THz, constructive interference is permitted along an angle of $\theta< 90^\circ$.

\section{Device fabrication and measurement setups}
The details of the epitaxial structure of the RTD oscillators are as those reported in Ref.\onlinecite{Han2023}. The peak current density of the RTD is 12~mA$/\mu m^2$, with a peak-to-valley current ratio (PVCR) of 2, and NDC region voltage swing of 0.7 V. A microscope of the fabricated arrays is shown in Fig.~\ref{fig:pic}(a). Twenty rows of the 1×11 slot arrays are fabricated within one region, allowing for selective biasing to each row independently. Further fabrication details can be found in the supplementary materials. The inset of Fig.~\ref{fig:pic}(a) provides a magnified view, with the common resistors and RTD areas labeled. The device scale aligns with the simulation model, featuring an L of 120~$\mu$m and an offset of 47~$\mu$m. Single RTD oscillators are also fabricated for comparison. 
We employ Fourier-transform infrared spectroscopy (FTIR) for frequency measurements. During the measurements, the RTD devices are positioned atop a Si lens, and biased using probes. For power measurement, a pyroelectric detector is utilized. Further details of the experimental setups can be found in the supplementary materials.
\section{Experiments}
\subsection{Even- and odd-mode emission from linear RTD oscillator arrays}
The red solid curve in Fig.~\ref{fig:pic}(b) shows a typical emission spectrum of a single RTD oscillator with a mesa size is 1~$\mu m^2$. The emission frequency is about 0.354~THz. The blue solid curve in Fig.~\ref{fig:pic}(b) shows a typical spectrum emitted from a row of RTD oscillators with a mesa size of about 0.9~$\mu m^2$.  There is only one peak in the emission spectrum and we interpret this as the evidence for coherent coupling of the RTDs in the row. Compared with the emission from a single RTD device, both the emitted power and frequency from the RTD array are significantly higher. 

\begin{figure}
  \includegraphics[width=0.8\linewidth]{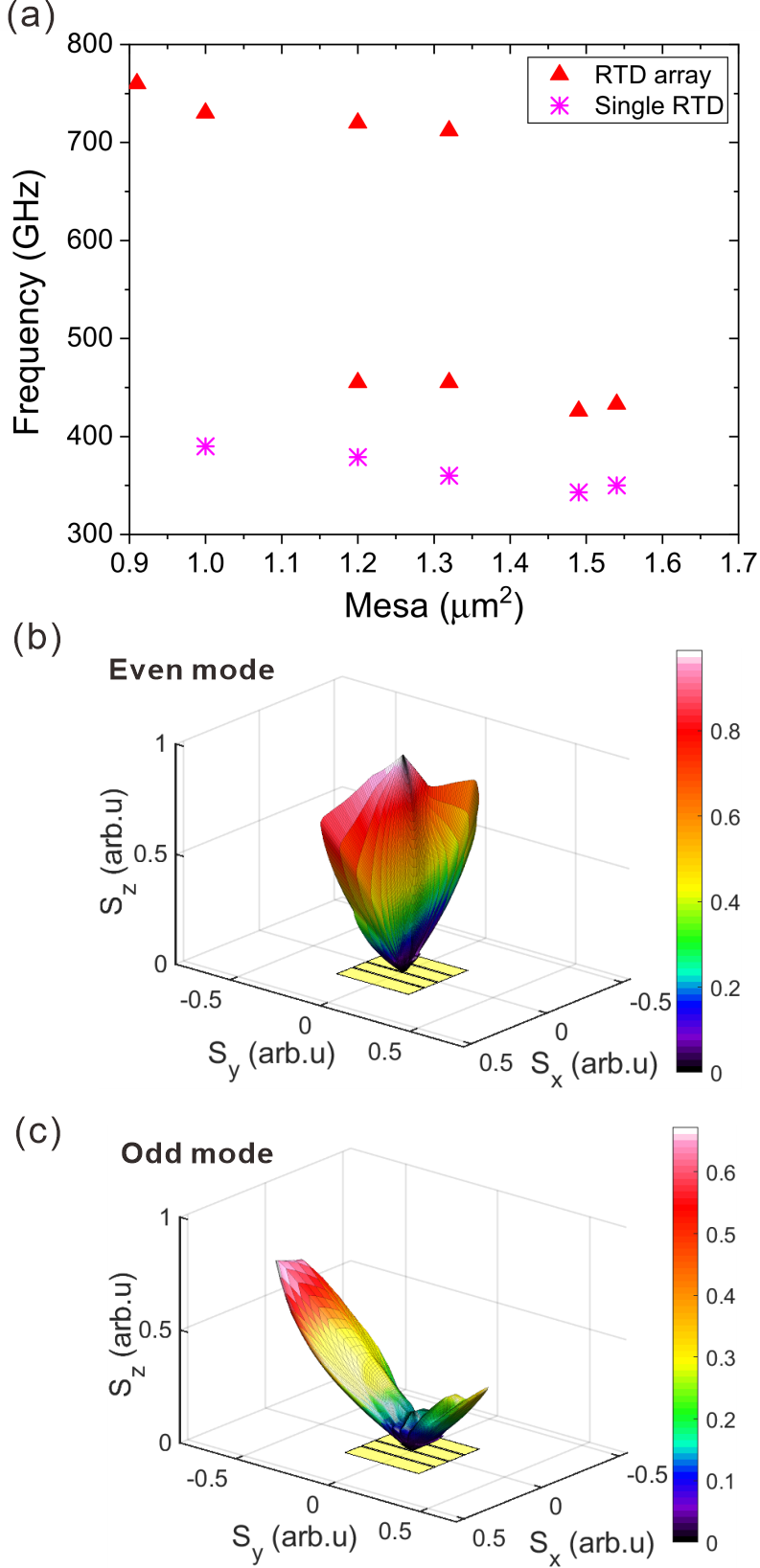}
  \caption{(a) Measured oscillation frequency of the single RTD oscillators and of the RTD oscillators in arrays, as a function of mesa area. The magenta double crosses represent the emissions from single RTD oscillators. The red triangles indicate emissions from RTD oscillators in arrays. (b, c) The measured far-field radiation pattern for the RTD oscillators in the array at mesa area of 1.33~$\mu$m$^2$, biased at a current of $~$0.8 A (b) and $~$0.75 A (c). The emissions point to the substrate side. S$_y$ indicates the direction along the row of 11-RTDs. S$_x$ indicates the direction perpendicular to the direction of S$_y$. }
  \label{fig:specM}
\end{figure}

Figure~\ref{fig:specM}(a) shows the emission frequency of the RTD oscillator as a function of the RTD mesa area. The magenta double crosses indicate the measurements from single RTD oscillators. The red triangles represent the emitted frequency from the RTD oscillators in the linear array. One notices that the single RTD oscillators emit radiation with frequencies below 0.4~THz. The emitted frequency from RTD oscillators in the arrays depends on the mesa areas: for small mesas, the RTD oscillators emit a high frequency at around 730 GHz, and for large mesas, the RTD oscillators emit a low frequency at around 730 GHz. The cross-over occurs at a mesa area of 1.3~$\mu$m$^2$, whose oscillation frequency can be switched by the applied current. We will discuss the switching of the oscillation modes in the later part.

The measured oscillation frequency as a function of mesa area is very similar to what we have predicted, as shown in Fig.~\ref{fig:structure}(b). The high-frequency emission corresponds to the even mode and the low-frequency emission corresponds to the odd mode. The measurements confirm the spectral distribution predicted from the modeling and simulations.

In order to evaluate the total emitted power, we measured the far-field emission patterns. We used a pyroelectric detector for the far-field measurements, the details of the measurements can be found in the supplementary materials. Fig.~\ref{fig:specM}(b) shows the far-field emission pattern for the array with a mesa area of 1.33~$\mu$m$^2$, when it was biased at high current/voltage (0.8~A) and emitted at high frequency. The main lobe of the emission is perpendicular to the sample surface pointing to the substrate side. This confirms the prediction of the even mode oscillation. Fig.~\ref{fig:specM}(c) shows the far-field emission pattern for the same array when it was biased at low current/voltage (0.75~A) and emitted at low frequency. The main lobe of the emission points at an angle of $\theta=48^\circ$ into the substrate side. By taking the frequency of 0.45~THz, we calculate the theoretical emission angle for the odd mode of about $\theta=53^\circ$. The theoretical prediction qualitatively fits with the measurements. The slight difference is attributed to the phases of RTD oscillators on both edges that do not have $\pi$ phase difference with neighboring slots, compared to that of other RTD oscillators in the middle of the array. The numerical simulations also reproduced the far-field patterns for both even and odd modes. The simulation results from CST software and their comparison with the measurements can be found in the supplementary materials.

 \begin{figure}
  \includegraphics[width=0.8\linewidth]{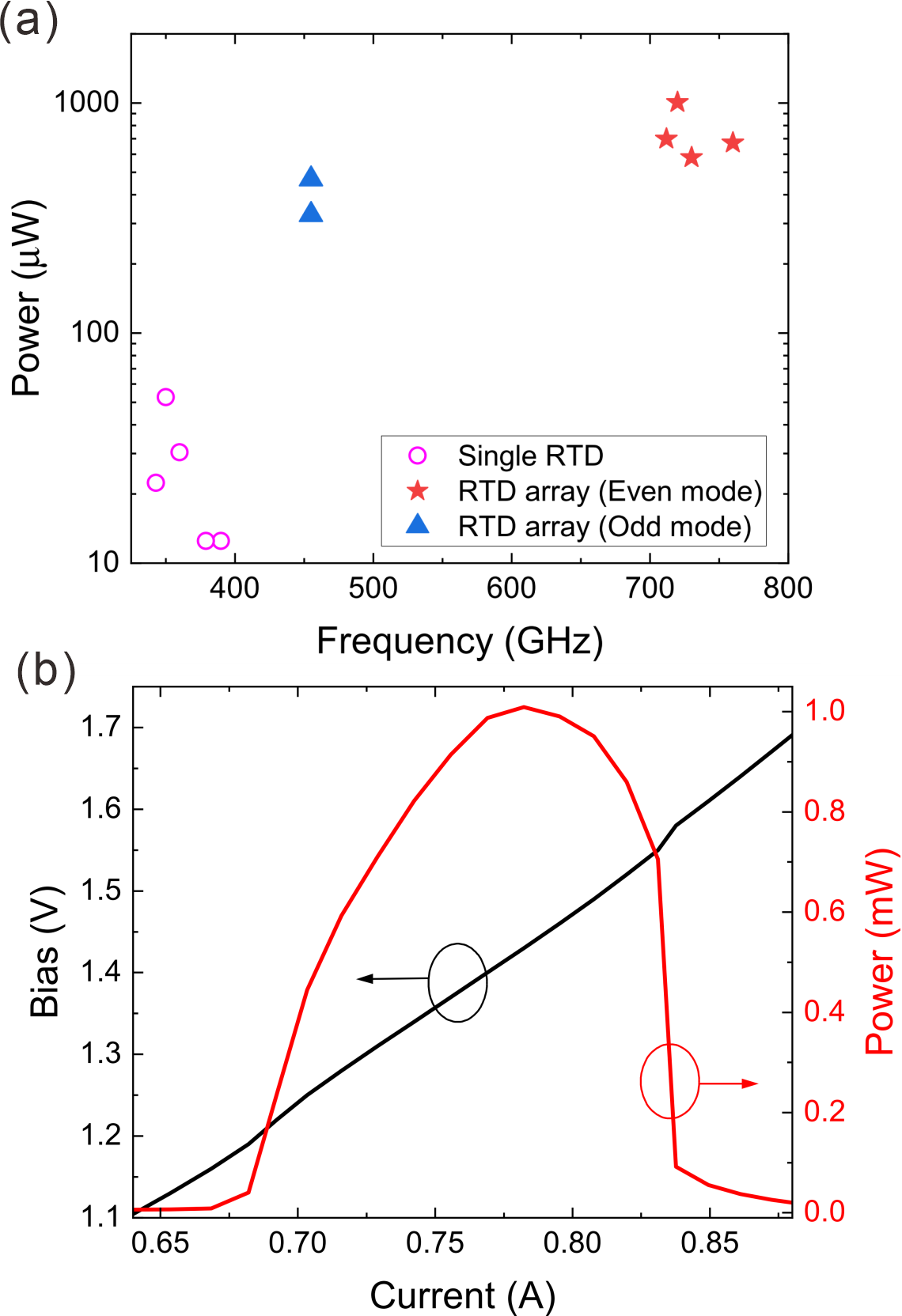}
  \caption{(a) Estimated output power of single RTD oscillators and oscillators in the arrays, as a function of oscillation frequency. The magenta circles represent the emissions from single RTD oscillators. The blue triangles denote the emissions from the odd mode oscillation of the arrays. The red stars indicate the emissions from the even mode of the arrays. (b) The power-current-voltage curves (L-I and I-V curves) for a typical linear array, with the mesa area of 1.2~$\mu$m$^2$. The black (red) curve represents the voltage (power) as a function of the applied current.}
  \label{fig:power}
\end{figure}

Based on the measured far-field pattern, we can estimate the total emitted power of RTD devices. Fig.~\ref{fig:power}(a) shows the estimated power as a function of emission frequency. For the single RTD oscillator, the emission power is in the range of several tens of $\mu$W. The emitted power from the 11-RTD-oscillator array has seen a significant increase: For the odd mode that oscillates at 0.45~THz, the power is about 400~$\mu$W; For the even mode that oscillates at 0.75~THz, the estimated power is about 1~mW. Fig.~\ref{fig:power}(b) shows the calibrated power-current-voltage (L-I and I-V) curves. From the I-V curve (black curve), one notices the NDC starts at 0.67 A (1.2 V) and ends at 0.83 A (1.6 V). The measured power also increases from 0.67 A and ceases at 0.83 A. Even though the current at higher voltage over the NDC region is still high, the detected power is weak, which indicates that the thermal emission is much weaker than the emission from the oscillation of RTDs. The thermal emission can be neglected. The estimated DC to AC conversion for the RTD oscillators in the array is about 0.08$\%$.

 \begin{figure}
  \includegraphics[width=0.8\linewidth]{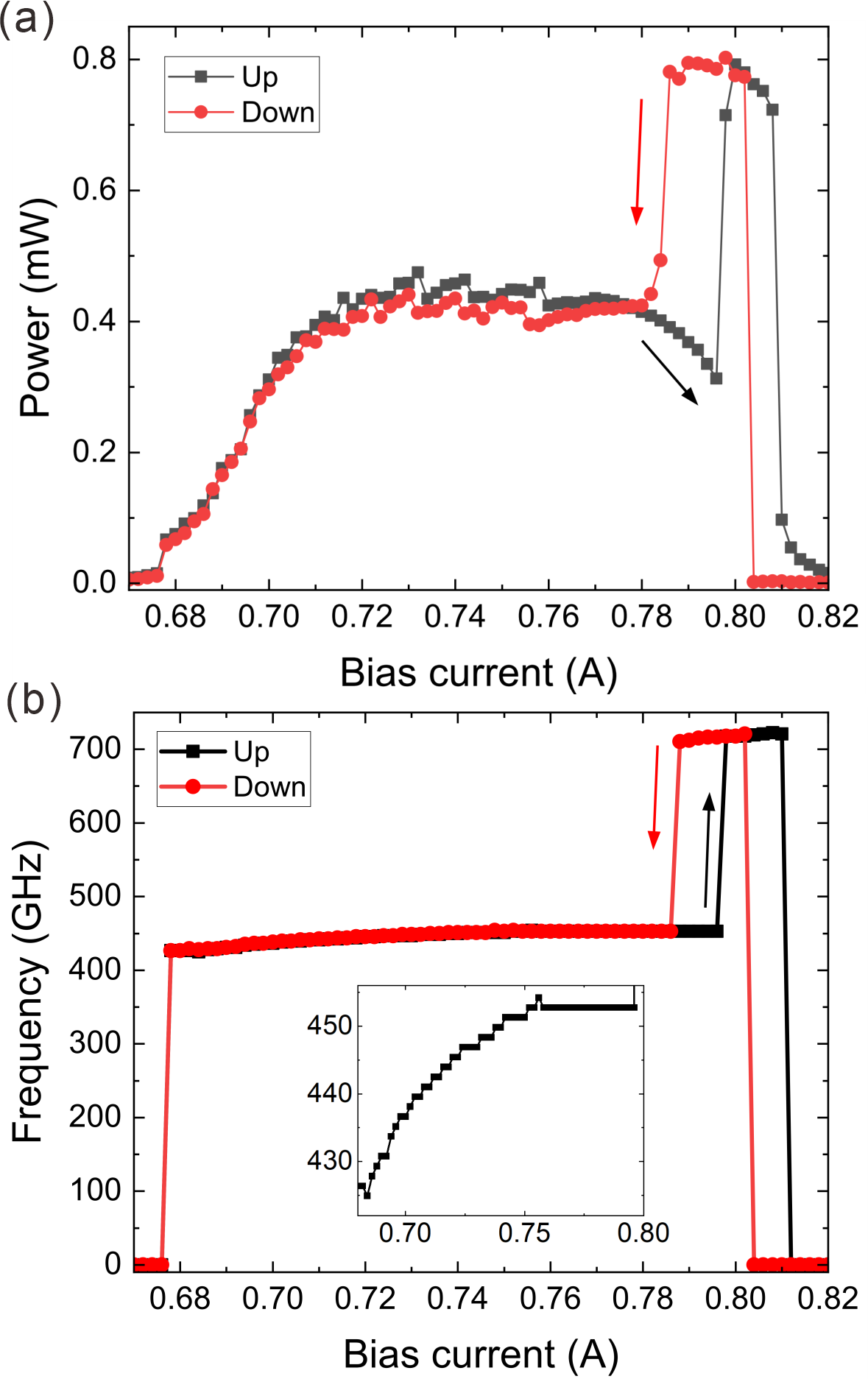}
  \caption{(a) The output power as a function bias current. The powers for the even and odd modes are calibrated. The black square connected line represents sweeping bias up. The red circle connected line represents sweeping bias down. (b) Measured frequency as a function of bias current. The black square connected line represents sweeping bias up. The red circle connected line represents sweeping bias down. The inset shows the enlarged frequency as a function of bias current (sweeping up). }
  \label{fig:dual}
\end{figure}

\subsection{Dual frequency band emission from RTD oscillator arrays}

As mentioned in the previous section when the mesa area is around 1.3~$\mu$m$^2$, the RTD oscillator array is switchable between two oscillation modes. Fig.~\ref{fig:dual}(a) and (b) show the measured power and frequency as a function of bias current for the linear RTD oscillator array with the mesa area of ~1.33 $\mu m^2$. The RTD array oscillate in two modes: when the bias current is low, the emitted power is low with the frequency lying at about 0.45~THz. The emission is attributed to the odd mode oscillation of RTDs. The inset of Fig.~\ref{fig:dual}(b) shows the enlarged oscillation frequency as a function of bias. The frequency can be tuned from 0.425~THz up to 0.452~THz, with a tuning range of 27 GHz. The maximum output power is about 450 $\mu$W. When the bias current is high, the emitted power is high (about 800 $\mu$W) with the frequency switched to about 0.71~THz. In this regime, the RTD array oscillates in the even mode, with a tuning range of 10 GHz. The frequency tuning of a single RTD device was reported\cite{Orihashi2005a}, and the increase of the oscillation frequency was attributed to the decrease of the capacitance of RTDs as the bias increases. In this particular RTD oscillator array, the substantial reduction in RTD capacitance effectively switches the oscillation mode from odd to even. Notably, this mode transition is accompanied by a directional shift in emission, showcasing the efficient beam steering capabilities of these dual-band operating RTD devices. 

Due to the nonlinear dynamics of RTD devices, the hysteresis features are expected at the start and end of the NDC region\cite{Feiginov2011}. The hysteresis in the current-voltage and power-voltage curves of the single RTD devices has been observed\cite{Ourednik2021}. In our RTD oscillator array, when we swept the direction of the bias, as indicated by the arrows in Fig.~\ref{fig:dual}(a) and (b) (the black arrow represents sweeping up and the red arrow indicates sweeping down), we observed pronounced hysteresis on the right side of the oscillation region. Interestingly, this strong hysteresis also appears at the bias point where the oscillation mode switches.

\section{Conclusion}
In conclusion, our study presents a novel structure for achieving high-power coherent emission in large-scale linear RTD-oscillator arrays. Particularly noteworthy is the ability of our RTD arrays to support both odd and even mode oscillations, with coherent emission observed in the far field for both modes. Furthermore, we successfully demonstrate dual-frequency band, mode-switchable emission from a single RTD array, accompanied by hysteresis performance during mode switching. Specifically, ~1 mW even-mode oscillation at ~750 GHz, and ~0.5 mW odd-mode oscillation at ~450 GHz are obtained with the fabricated mode-switchable 1×11 array.

These findings highlight the versatility and controllability of our RTD array platform, offering exciting opportunities for the development of advanced~THz emitters. By leveraging the unique properties of RTD-oscillator arrays, including their coherent emission and dual-frequency band capabilities, our study contributes to the advancement of THz technology and opens new avenues for applications in communication, sensing, and imaging.

Future research efforts may focus on further optimizing the design and performance of RTD-oscillator arrays, exploring additional functionalities, and advancing integration techniques for enhanced practical utility in various THz applications. Overall, our work underscores the promising potential of RTD-oscillator arrays as key components in the realization of high-performance THz systems.

\begin{acknowledgments}
FM, JH, and HGR acknowledge the financial support from DFG projects RO 770/46-1 and RO 770/50-1 (the latter being part of the  DFG-Schwerpunkt "Integrierte Terahertz-Systeme mit neuartiger Funktionalität (INTEREST)"). PO and MF thank the financial support from FWF project P30892-N30. TZ and SS acknowledge a scientific grant-in-aid (24H00031) from JSPS and CREST (JPMJCR21C4) from JST.
\end{acknowledgments}

\section*{AUTHOR DECLARATIONS}
\subsection*{Conflict of Interest}
The authors have no conflicts to disclose.


\section*{DATA AVAILABILITY}
The data that support the findings of this study are available from the corresponding authors upon reasonable request.

\bibliography{library}

\end{document}